\documentclass[aps,prl,amsmath,twocolumn]{revtex4-2}

\usepackage{graphicx}
\usepackage{dcolumn}
\usepackage{bm}
\usepackage{amsmath}
\usepackage{float}
\usepackage{color}

\begin{document}


\title{Bandwidth bounds for wide-field-of-view dispersion-engineered achromatic metalenses}

\date{June 15, 2021}

\author{Kunal Shastri}
\email[]{kks75@cornell.edu}
\affiliation{School of Electrical and Computer Engineering, Cornell University, Ithaca, NY 14853, USA}

\author{Francesco Monticone}
\email[]{francesco.monticone@cornell.edu}
\affiliation{School of Electrical and Computer Engineering, Cornell University, Ithaca, NY 14853, USA}

\begin{abstract}
Optical systems with wide field-of-views (FOV) are crucial for many applications such as high performance imaging, optical projection, augmented/virtual reality, and miniaturized medical imaging tools. Typically, aberration-free imaging with a wide FOV is achieved by stacking multiple refractive lenses (as in a ``fisheye'' lens), adding to the size and weight of the optical system. Single metalenses designed to have a wide FOV have the potential to replace these bulky imaging systems and, moreover, they may be dispersion engineered for spectrally broadband operation. In this paper, we derive a fundamental bound on the spectral bandwidth of dispersion-engineered wide-FOV achromatic metalenses. We show that for metalenses with a relatively large numerical aperture (NA), there is a tradeoff between the maximum achievable bandwidth and the FOV; interestingly, however, the bandwidth reduction saturates beyond a certain FOV that depends on the NA of the metalens. These findings may provide important information and insights for the design of future wide-FOV achromatic flat lenses.  	

\end{abstract}

\maketitle


\section{Introduction}  

For more than the last three hundred years, optical systems have mostly relied on refractive elements, in the form of lenses for focusing/imaging applications. From a wave optics perspective, these lenses are based on the principle that a continuous phase profile can be added to the incident wavefront through propagation in a dielectric material of inhomogenous thickness. Although dielectric diffractive lenses have a flatter and thinner profile than conventional lenses, as they introduce phase modulo $2\pi$, this phase is still obtained in the same way as in a refractive lens, through wave propagation in a dielectric material of varying thickness \cite{sweeney_harmonic_1995,engelberg_advantages_2020, banerji_imaging_2019,engelberg_achromatic_2020}. 

In recent decades, new intriguing opportunities in lens design have emerged with the introduction of metasurfaces, flat and planar optical components that can abruptly change the phase of the incident wavefront over the distance of a wavelength or less \cite{ding_gradient_2018, genevet_recent_2017}. This is achieved not (or not only) through wave propagation inside a dielectric material, but through wave interaction with arrays of sub-wavelength nano-structures, called meta-atoms, which scatter light with the desired phase shift, allowing for a local control of the wavefront with great flexibility \cite{decker_imaging_2019, arbabi_subwavelength-thick_2015}. Lenses designed using metasurfaces, called metalenses, offer flat, compact and lightweight alternatives to conventional lenses, features that are especially important in consumer electronics such as high-performance smartphone cameras and augmented/virtual reality glasses \cite{lalanne_metalenses_2017,tseng_metalenses_2018, khorasaninejad_metalenses_2017}. By engineering the response of complex nanostructures used as meta-atoms, metalenses have also been shown to enable additional functionalities, such as polarization control, that cannot be easily achieved using conventional refractive and diffractive lenses \cite{yu_flat_2014, yu_polarization-independent_2020}.   
 
\begin{figure}
 	\includegraphics[width=\columnwidth]{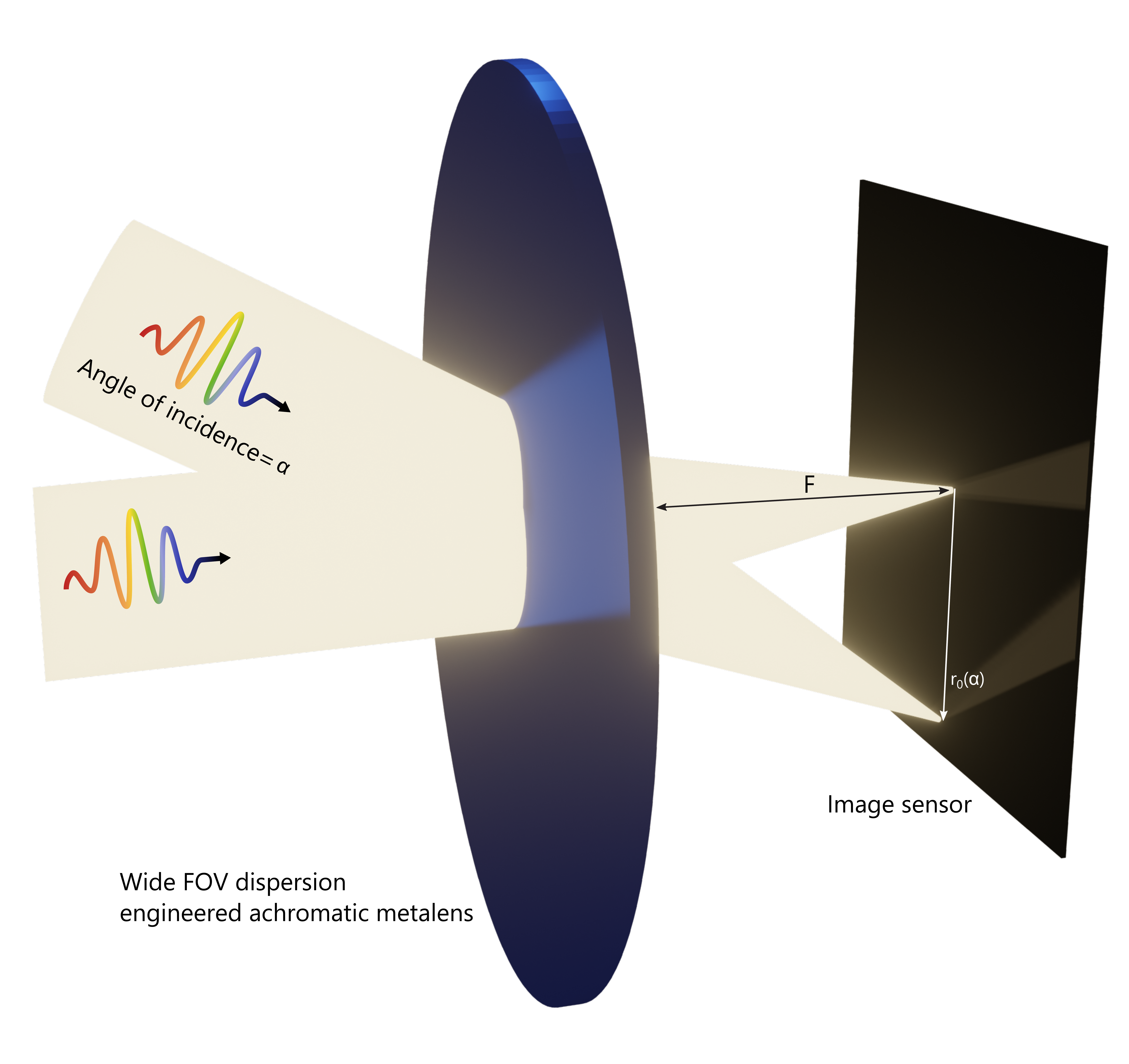}
 	\caption{Schematic of an achromatic metalens with a wide field-of-view. The metalens should be able to transform the planar wavefront incident at any given angle into a spherical wavefront that converges at a point on the focal plane. The location of the focal spot, indicated as $r_0$, necessarily depends on the angle of incidence of the plane wave, but, for achromatic operation, it should not depend on frequency.} \label{fig:1}
\end{figure}

Unfortunately, the abrupt phase shift introduced by metasurfaces is typically strongly dependent on the wavelength of the incident light making it challenging to design broadband metalenses. An elegant strategy to overcome this limitation is to develop a library of meta-units that provide, not only the desired phase pattern, but also the appropriate group delay and group delay-dispersion, as recently proposed in Refs. \cite{chen_flat_2020, arbabi_controlling_2017, shrestha_broadband_2018, chen_broadband_2018, chen_broadband_2019}. This can be illustrated by considering a metasurface that changes the phase of an incident plane wave to a generic phase profile given by $\phi(r,\omega)=-\zeta(r)\omega/c$, where $\omega$ and $c$ are the angular frequency and speed of light in the surrounding medium, respectively, and the function $\zeta(r)$ has units of length. For example, for a linear phase gradient that redirects an incident beam at an angle $\alpha$, we have $\zeta(r)=r\sin(\alpha)$, and, for a metalens with focal length $F$ that focuses normally incident light, $\zeta(r)=(F^2+r^2)^{1/2}-F$, where $r$ is the in-plane linear/radial coordinate, respectively. For an ideal broadband achromatic metasurface, the meta-atoms at each location on the metasurface must be chosen to perfectly satisfy the above phase requirement at every frequency. An alternative, equivalent way to interpret this requirement is that the meta-atoms must be chosen to satisfy the required phase profile at the center frequency $\omega_c$, along with a constant group-delay profile given by $t_g(r)=\partial\phi(r,\omega)/\partial\omega|_{\omega_c}=-\zeta(r)/c$, and a group-delay dispersion equal to zero (achromatic wavefront manipulation, for example achromatic focusing or deflection, would still be obtained even if the group delay dispersion is non-zero, as long as it remains spatially constant, but pulses would be distorted). Metalenses designed using this method, implementing the required phase and group-delay profile, are called dispersion-engineered achromatic metalenses, and they can correct for chromatic aberrations over a continuous frequency range centered around the chosen center frequency, as originally demonstrated in \cite{shrestha_broadband_2018}. 

Recently, by drawing parallels between a dispersion-engineered achromatic metalens and a series of delay lines and phase shifters, it was shown in Ref. \cite{presutti_focusing_2020} that the range of group delays that can be implemented by a meta-atom cannot be arbitrary but is bound by delay-bandwidth limitations (the product between group delay and signal bandwidth is fundamentally limited in any linear time-invariant system \cite{miller_fundamental_2007-1,miller_fundamental_2007}). Consequently, this places a theoretical upper bound on the bandwidth of the dispersion-engineered metalens, regardless of the chosen meta-atom library. In this paper, we generalize the bandwidth bounds to wide field-of-view (FOV) \cite{huang_achromatic_2021, lin_computational_2021, martins_metalenses_2020, fan_ultrawide-angle_2020, chen_-chip_2020, guo_high-efficiency_2018} achromatic metalenses. In particular, we show that the maximum achievable bandwidth for dispersion-engineered metalenses decreases not only with an increase in numerical aperture (NA) \cite{presutti_focusing_2020}, but also with an increase in the FOV.

\begin{figure*}
 	\includegraphics[width=0.9\textwidth]{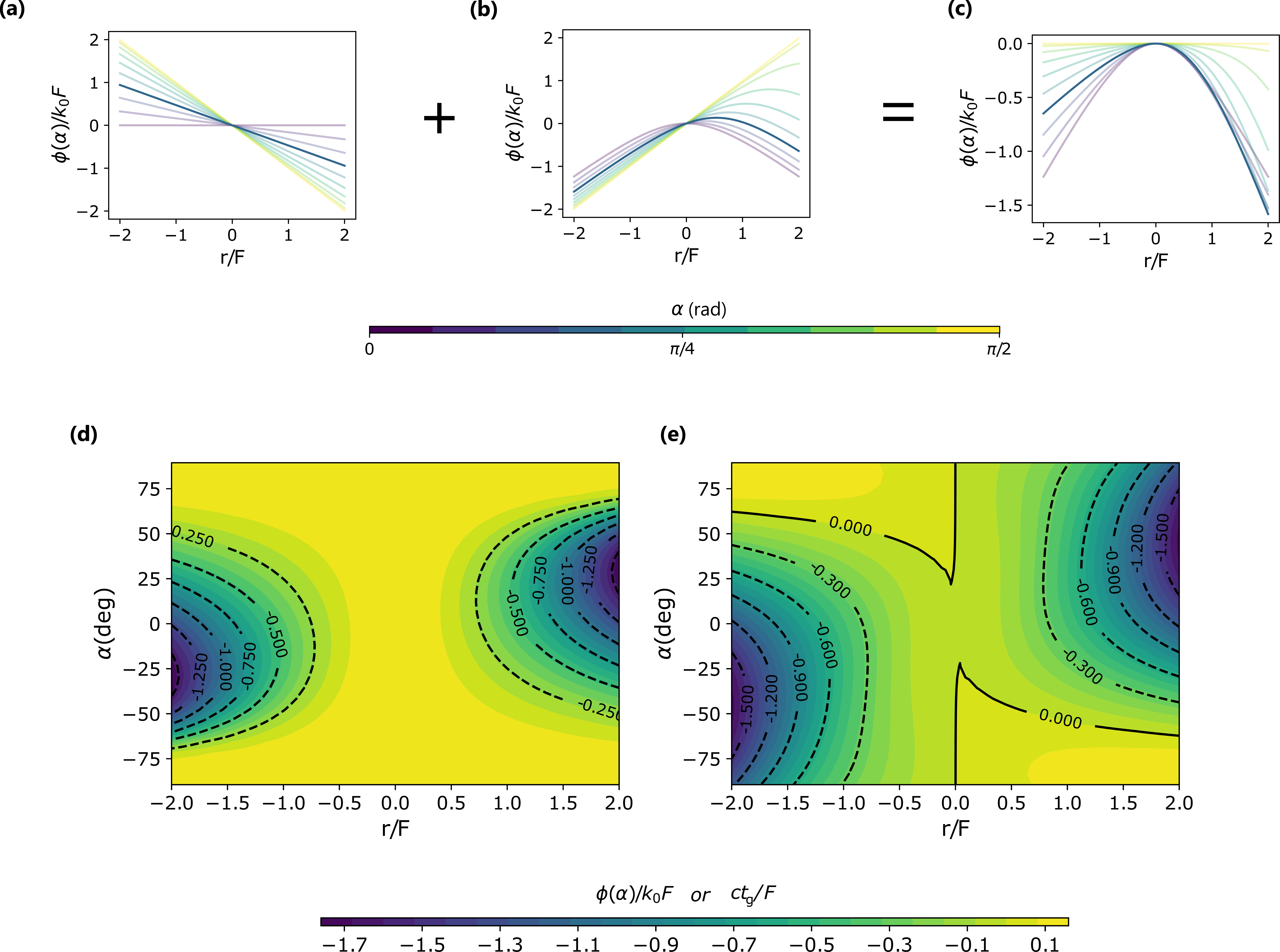}
 	\caption{Ideal spatial phase profile for focusing an obliquely incident plane wave on a planar focal plane at a distance $F$ from a planar metalens. The ideal phase profile for any given angle of incidence can be expressed as the sum of a linear phase gradient (a) to redirect the wave to the surface normal and a hyperbolic phase profile (b) to focus the wave. In (a)-(c) phase profiles for different polar angles of incidence, $\alpha$, are represented using different colors and $\alpha=0$ corresponds to normal incidence. One particular phase profile for angle of incidence equal to 30 deg is highlighted with a bolder line. (d), (e) Ideal phase profile, as a function of the angle of incidence, $\alpha$, for two different choices for the location of the focal spot: (d) $r_0=F\tan(\alpha)$ and (e) $r_0=F\alpha$. The phase is normalized by the product of the wavenumber, $k_0=\omega/c$, and the focal length of the metalens, $F$, so that both plots can also be interpreted as the normalized group delay ($ct_g/F = c~ \partial\phi(r,\omega)/\partial\omega|_{\omega_c}/F$) required for achromatic focusing. In all plots `$r$' is the radial coordinate on the metalens. } \label{fig:2}
\end{figure*}

\section{Results}  
 
\textit{Location of the focal spot --} One challenge in the design of wide-FOV metalenses is that plane waves incident on the lens at various angles need to be phase shifted by different values in order to avoid aberrations. In particular, a metalens with focal length $F$ that changes the phase of an incoming plane wave to a hyperbolic profile given by,
\begin{equation} \label{eq:phase1}
\phi(r,\omega)=-\frac{\omega}{c}\left(\sqrt{F^2+r^2}-F\right),
\end{equation}
achieves aberration-free focusing only for normally incident light \cite{aieta_aberrations_2013}. While some workarounds for this problem exist, such as using a curved (aplanatic) metalens \cite{aieta_aberrations_2013} or a curved focal surface \cite{hunt_broadband_2011}, these devices are not ``flat", negating some of the potential benefits of using metalenses. Thus, in this work we focus on flat-optic systems consisting of a planar metalens and a planar focal surface, as sketched in Fig \ref{fig:1}. In this case, while normally incident rays are focused at the center of the focal plane, to create an aberration-free image over a wide FOV, obliquely incident rays need to be focused on the same plane at a different location $r_0(\alpha)$ relative to the center. Moreover, to create an undistorted image of the considered scene, it is necessary that $r_0(\alpha)=F\tan(\alpha)$, where $\alpha$ is the angle of incidence and $F$ is the focal length of the metalens \cite{kalvach_aberration-free_2016}. Note that since the problem is cylindrically symmetric, it is sufficient to use a single polar angle $\alpha$, with $\alpha=0$ corresponding to normal incidence, to describe the phase shift. Since this mapping from angle of incidence to focal spot position requires an infinitely large sensor for grazing angles of incidence, another frequently used choice is $r_0(\alpha)=F\alpha$, which however introduces significant distortions for large incident angles. 

\textit{Ideal metalens phase profile --} To focus a planar wavefront incident at a given angle $\alpha$ on a flat focal plane, the lens must impart a combination of a linear phase gradient given by,
\begin{equation} \label{eq:phase2}
\phi_l(r,\omega,\alpha)=-\frac{\omega}{c}\left(r\sin(\alpha)\right),
\end{equation}
to redirect the wavefront to the surface normal, and the hyperbolic radial profile given by Eq. (\ref{eq:phase1}) but focusing light at $r_0$ instead of the center of the plane \cite{shalaginov_single-element_2020},
\begin{equation} \label{eq:phase3}
\begin{split}
\phi_s(r,\omega,\alpha)&=-\frac{\omega}{c}\left(\sqrt{F^2+(r-r_0(\alpha))^2}\right.\\
&\left.-\sqrt{F^2+r_0(\alpha)^2} \right), 
\end{split}
\end{equation}
Thus, the phase shift that a wide-FOV metalens must introduce for focusing without aberrations is,
\begin{equation} \label{eq:phasef}
 \phi(r,\omega,\alpha)=\phi_l(r,\omega,\alpha)+\phi_s(r,\omega,\alpha). 
\end{equation}

\begin{figure*}
	\includegraphics[width=0.85\textwidth]{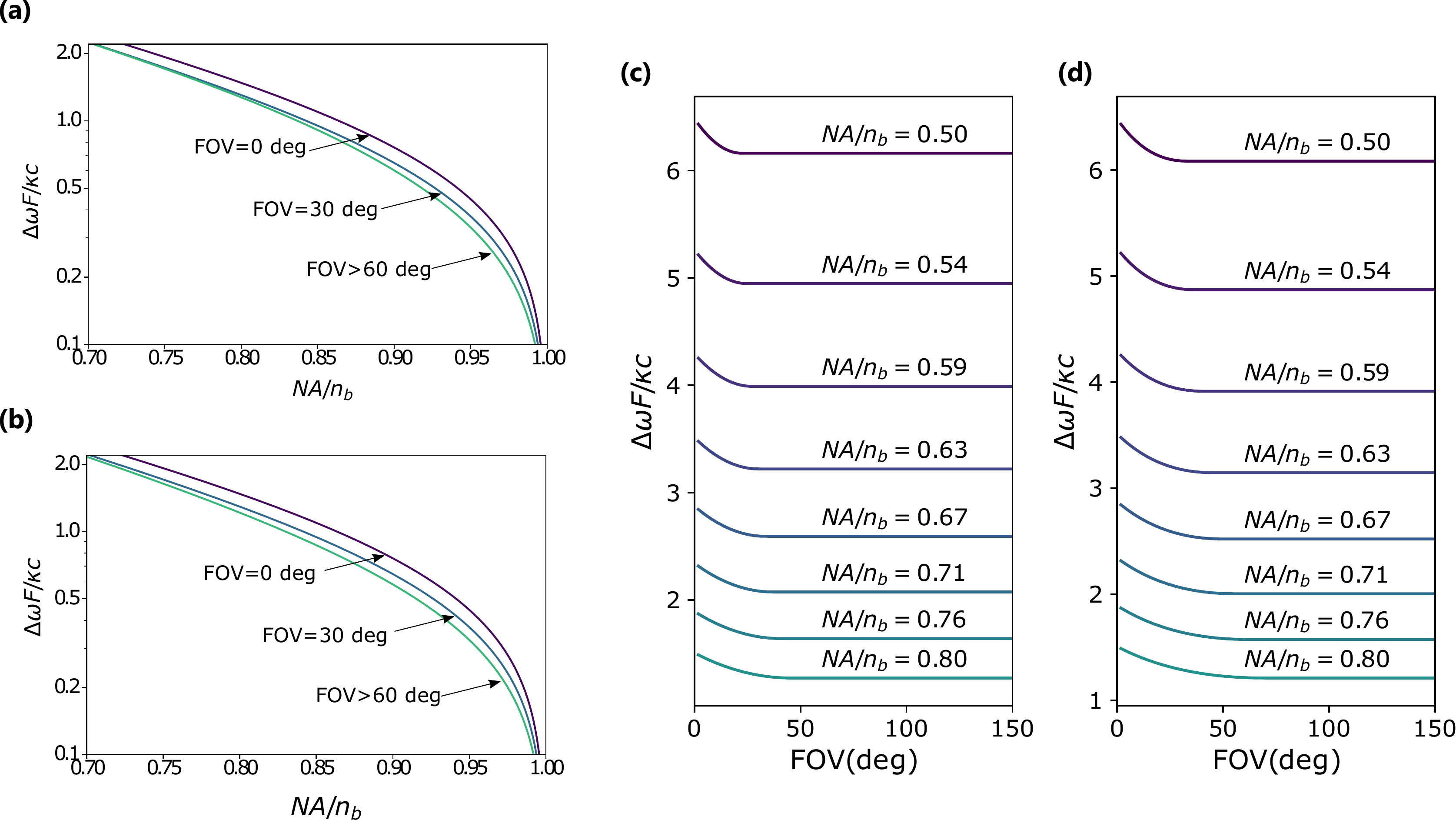}
	\caption{Calculated bandwidth bounds for wide-FOV dispersion-engineered achromatic metalenses. In (a) and (c), the bound is calculated for a focal spot location given by $r_0=F\tan(\alpha)$ and in (b) and (d) for $r_0=F\alpha$. To represent the bound in a general way applicable to vastly different metalenses, the bandwidth is normalized by the factor $\kappa$, the focal length of the metalens $F$, and the speed of light $c$. The bandwidth axis in (a) and (b) has been scaled logarithmically for clarity. In all the panels, $NA$, $n_b$, and FOV are the numerical aperture, background refractive index, and the field of view, respectively.} \label{fig:3}
\end{figure*}

The spatial dependence of the two individual phase profiles and the resultant net phase shift for $r_0(\alpha)=F\tan(\alpha)$ are plotted in Figs. \ref{fig:2} (a)-(c). Furthermore, the calculated net phase profiles for two different choices of $r_0(\alpha)$ are shown as a function of the angle of incidence in Figs. \ref{fig:2} (d),(e). Importantly, it can be seen that the ideal phase profile has a strong dependence on the direction of the incident light and, for a metalens of finite lateral extent, the required range of phase shifts does not increase monotonically with the angle of incidence. In fact, while the required total phase range would be the same regardless of incident angle for an ideal, infinitely extended metalens, for a finite metalens the phase range to be imparted is minimal for grazing angles of incidence, as seen in Figs. \ref{fig:2} (c),(d), and is maximum at a certain angle that depends on the maximum radius of the metalens.       

An important point to note is that a ``local" metasurface can only introduce a position-dependent phase shift and not an angle-dependent shift. Hence, while a local metasurface can be used to focus light incident at a specified angle, using the angle-appropriate phase profile above, it will form a distorted and aberrated image for light incident at other angles. On the other hand, a lens with a FOV equal to $2\alpha_0$ should be able to focus light from all angles between $-\alpha_0$ and $\alpha_0$ without any aberrations. 

The easiest way to achieve an angle-dependent phase shift with local metasurfaces is to consider a system with at least two surfaces separated by a gap. The gap here is crucial since it utilizes the angle-dependent phase accumulated by plane waves propagating through a free-space (or dielectric) slab to create an overall angle-dependent transmission phase. In practice, this has been implemented using two metasurfaces separated by a certain distance \cite{huang_achromatic_2021, arbabi_miniature_2016, groever_meta-lens_2017} or a single metasurface preceded by a circular aperture \cite{kalvach_aberration-free_2016, shalaginov_single-element_2020, fan_ultrawide-angle_2020, guo_high-efficiency_2018, yang_design_2021}. The aperture followed by the free-space gap maps incident beams from different angles onto separate (or minimally overlapping) spatial regions on the local metasurface, which can then implement the necessary spatial phase shift (a nice example of this design strategy is demonstrated in Ref. \cite{shalaginov_single-element_2020}). In general, however, the bandwidth limitations for dispersion-engineered achromatic metalenses that will be derived in the next section are not affected to the presence of the aperture, since the required range of group delays remains unchanged. 

To achieve an angle-dependent phase shift with a single metasurface, it would instead be necessary to design angle-dependent meta-atoms providing different phase shifts for different angles \cite{kamali_angle-multiplexed_2017} or, more broadly, ``nonlocal metasurfaces'' with an overall angle-dependent response \cite{silva_performing_2014,monticone2016parity,overvig_thermal_2021,overvig_multifunctional_2020,reshef2021optic,guo2020squeeze,chen2021dielectric,shastri2022extent}. As further discussed in the Discussion section, nonlocal metasurfaces may provide some intriguing opportunities to overcome the limitations discussed in the following, which are based on the assumption of local interactions between incident wave and meta-atoms.

\textit{Maximum achievable bandwidth --} Next, we proceed to calculate the fundamental bandwidth bounds for wide-FOV metalenses that have been designed, through dispersion engineering, for broadband achromatic operation (i.e., for diffraction-limited focusing at the same distance over a broad frequency range). As discussed in the Introduction, the meta-atoms composing these structures are designed to implement not only  the phase shift given by Eq. (\ref{eq:phasef}) at the center frequency, but also the required group delay for broadband focusing. The group delay required for a pulse incident on the wide-FOV metalens at a location $r$ relative to the center of the lens, and at an angle $\alpha$,  can be calculate by taking the angular frequency derivative of Eq. (\ref{eq:phasef}), 

\begin{equation} \label{eq:delt}
\begin{split}
t_g(r,\alpha)&=-\frac{F}{c}\left[\frac{r}{F}\sin(\alpha)+\sqrt{1+\left(\frac{r}{F}-\frac{r_0(\alpha)}{F}\right)^2} \right.\\
&\left.-\sqrt{1+\frac{r_0(\alpha)}{F}^2}\right].
\end{split} 
\end{equation}

In this equation the relative group delay is negative at any position $r\neq0$, which can be interpreted as a time advance for a pulse at the edges of the metalens with respect to an undelayed pulse at the center, or as a (more physical) time delay at the center with respect to an undelayed pulse at the edges. Importantly, adding a global (i.e., position-independent) time delay is irrelevant for our purposes. Note that the plots in Figs. \ref{fig:2}(d) and (e) can also be used as representations of the group delay profile. This is because the spatial phase profile is proportional to the group delay profile (as can be seen by comparing Eqs. (\ref{eq:phasef}) and (\ref{eq:delt})) and they can be made numerically equal using an appropriate choice of normalization. The plot in Fig. \ref{fig:2}(d) corresponds to the group delay calculated for a choice of focal spot location $r_0=F\tan(\alpha)$, and (e) for $r_0=F\alpha$. 

Since the thickness of typical metalenses is of the order of the wavelength, this group delay is primarily acquired during longitudinal propagation through the metalens  (namely, delay acquired through lateral propagation can be neglected, as in \cite{presutti_focusing_2020}). Thus, each point on the metalens can be modeled as a 1D delay line designed to implement the local delay given by Eq. (\ref{eq:delt}). This group delay is typically introduced by using resonant meta-atoms or, alternatively, meta-atoms acting as truncated waveguide segments with engineered dispersion. In the case of a structure supporting a single resonance, a simple study of the dynamics of the mode amplitude shows that the total rate of loss $\gamma$, due to absorption and radiation, determines both the full-width-at-half-maximum bandwidth of the resonance $\Delta \omega = 2\gamma$ and the mode lifetime $\tau=1/\gamma$, the latter of which corresponds to the maximum delay achievable through the interaction with the cavity. The delay-bandwidth product is therefore, $\Delta T\Delta\omega=2$, for any single-mode resonant cavity \cite{mann_nonreciprocal_2019} and it can be easily verified that, for multi-resonator systems, the product scales with the number of resonators \cite{wang2003compact,tucker_slow-light_2005}. Furthermore, more general limits on the delay-bandwidth product can be derived for generic 1D systems, applicable also to non-resonant waveguiding structures. A particularly general expression for this limit was derived by Miller in \cite{miller_fundamental_2007, miller_fundamental_2007-1}: $\Delta T\Delta\omega\leq \kappa = L\omega_c\eta/2\sqrt{3}c$, where $L$ is the length of the structure, $\omega_c$ is the central frequency, and $\eta$ is the maximum contrast in relative permittivity at any frequency within the considered bandwidth and any point within the structure. While the derivation of this result is rather involved, the bound can be intuitively understood, in the case of dielectric waveguides, considering the dispersion diagram (band diagram) of the waveguide modes: since the guided-mode bands are confined between the light lines defined by the permittivities of the dielectric waveguide and the surrounding material, reducing the group velocity (i.e., increasing the group delay) by flattening a band between these two lines necessarily implies a reduction of the bandwidth over which the desired group velocity can be achieved. Applied to our case of interest, these bounds on the delay-bandwidth product allow limiting the maximum bandwidth over which a meta-atom is able to implement a certain group delay, and, in turn, the maximum operating bandwidth of the entire dispersion-engineered achromatic metalens. For further details on the relevance of the delay-bandwidth product for metalenses and metasurfaces, see Refs. \cite{presutti_focusing_2020, engelberg_achromatic_2020,shastri2022extent}.

Based on the above considerations, the maximum frequency range for achromatic operation for a metalens with a FOV equal to $2\alpha_0$ can then be calculated by finding from Eq. (\ref{eq:delt}) the maximum required group delay within the angular range $-\alpha_0\leq\alpha\leq\alpha_0$, and substituting it in the delay-bandwidth bound,

\begin{equation} \label{eq:bandwidth}
\Delta\omega\leq\frac{\kappa}{\Delta T_{\max}}. 
\end{equation}

The group delay range that the metalens should provide widens for larger lenses, as larger delays need to be implemented at the center ($r=0$) to compensate for the additional time taken by a pulse arriving to the focal spot from the edges ($r=R$). The required delay range also widens for larger FOV (wider range of angles of incidence), since, intuitively, an even greater delay needs to be compensated across the metalens as the angle of incidence increases. However, this is true only up to a point. Indeed, the range of required group delays across a finite metalens does not monotonically widens as the angle of incidence increases owing to the non-monotonic increase of phase range with angle that we discussed in the previous section (see Figs. \ref{fig:2}(d),(e)). Taking this into account, the greatest overall delay that the metasurface must be able to implement can be calculated as $\Delta T_{\max}=t_g(r=0,\alpha=0)-t_g(r=R,\alpha=\alpha^*)$, where $\alpha^*$ is the angle, within the range $-\alpha_0\leq\alpha\leq\alpha_0$, for which the group-delay range across the metalens is widest (essentially this is the difference between the maximum and the minimum in the contour plots in Fig. \ref{fig:2}(d,e), for any angle $\alpha$ and position $r$, when these plots are interpreted as a representation of the normalized delay). Since there is no simple expression for $\Delta T_{\max}$, we evaluate it numerically and then use it in Eq. (\ref{eq:bandwidth}) to determine the bandwidth limit for two different choices of focal position $r_0$. The results are plotted in Fig. \ref{fig:3}.

As a function of the numerical aperture (NA) of the metalens, Figs. \ref{fig:3}(a) and (b) show a clear trend of narrower achievable bandwidths for metalenses with larger NA, which is expected, since a metalens with larger NA has a larger delay to compensate between the center and the edges, consistent with \cite{presutti_focusing_2020}. Here, we only plotted the results for moderately large NA because, for smaller values, the bandwidth bound for any FOV approaches the normal-incidence bound derived in \cite{presutti_focusing_2020}. Most importantly, it can also be seen from these plots that the bandwidth limit shrinks for metalenses with larger FOV. However, surprisingly, this trend saturates for a certain FOV, depending on the NA and the choice of focal spot position, as seen in Figs. \ref{fig:3}(c) and (d). This implies, for example, that a  dispersion engineered metalens with a ultra wide FOV=180 deg could, in principle, have the same spectral bandwidth as a metalens with a relatively narrower FOV=60 deg, for a fixed $NA=0.7n_b$. This result is a consequence of the previously mentioned observation that for a finite metalens and a fixed FOV, the widest group-delay range that the metalens should implement is not always for light incident at the maximum oblique angle, implying that beyond a certain FOV (dependent on NA), the maximum allowed bandwidth does not further decrease. Furthermore, this trend is observed for both choices of focal spot position, as seen in Figs. \ref{fig:3}(c) and (d).

\section{Discussion}  

In the recent literature, moderately broadband achromatic metalenses have been reported with remarkably large FOVs, as well as sufficiently low aberrations resulting in a high Strehl ratio \cite{kalvach_aberration-free_2016, shalaginov_single-element_2020, fan_ultrawide-angle_2020, guo_high-efficiency_2018, yang_design_2021}. In particular, a complete 180 deg FOV over a relatively wide fractional bandwidth of 0.2 has been reported in \cite{yang_design_2021}. However, the NA in this design is limited to $0.24$ for which our bounds are less relevant since, as seen in Figs. \ref{fig:3}(a) and (b), an appreciable decrease in bandwidth as a function of FOV is only seen for metalenses with a relatively large NA. In this context, it would be interesting to see how future wide-FOV metalens designs with larger NA compare with our bounds. 

In relation to the bandwidth bounds derived above, it is also instructive to consider structures that can potentially break some of the assumptions in the derivation and thereby overcome the bound. Indeed, the bandwidth bounds derived here apply to the spatial phase profile given by Eq. (\ref{eq:phasef}), which is specific to the case of diffraction-limited focusing on a flat focal plane. For other imaging system designs, such as focusing on curved focal surfaces, the resultant bound will have to be modified depending on the specific spatial phase profile of the considered metalens and the desired imaging functionality. In addition, relaxing the constraint on diffraction-limited resolution and Strehl ratio would imply that a certain level of error in the implemented phase/group delay profile would be acceptable, which would therefore relax the resulting bandwidth bounds, as also demonstrated in Ref. \cite{presutti_focusing_2020}.

Moreover, since the considered limits to the delay-bandwidth product strictly only apply to 1D structures (the meta-atoms modeled as 1D delay lines), the derived bound will not apply to systems that allow additional group delay due to transverse propagation. For instance, systems consisting of cascaded metasurfaces could take advantage of the lateral dimension between metasurfaces to achieve larger group delays and potentially overcome the bandwidth bounds derived in this paper (however, a generalized version of these bounds should apply also to this type of systems, and its derivation will be the subject of future work). The same argument also applies to wide-FOV imaging systems constructed using bulk metamaterials, or thick volumetric structures obtained, for example, through all-area inverse design \cite{lin_computational_2021}. More broadly, we note that the analogy between delay lines and meta-atoms is valid only if the metasurface response is based on the \emph{local} interaction between fields and meta-atoms at a specific point on the surface. Differently from achromatic metalenses, achromatic diffractive lenses \cite{banerji_imaging_2019} achieve broadband operation through energy redistribution among different diffraction orders for different frequencies, a mechanism that involves the overall surface structure rather a local effect, as discussed in \cite{engelberg_achromatic_2020}. For this reason, achromatic diffractive lenses, unlike dispersion-engineered metalenses, are not limited by delay-bandwidth considerations, but suffer from other limitations especially in terms of focusing power \cite{engelberg_achromatic_2020}. This general discussion also hints at the large potential of nonlocal metasurfaces \cite{silva_performing_2014,monticone2016parity,overvig_thermal_2021,overvig_multifunctional_2020,reshef2021optic,guo2020squeeze,chen2021dielectric,shastri2022extent} -- an emerging class of metasurfaces specifically designed to exhibit strong nonlocal effects -- to potentially achieve performance metrics that are unattainable with more conventional designs.

To conclude, in this paper we have generalized our previous work on metalens bandwidth bounds to dispersion-engineered achromatic metalenses with a wide FOV. Single element (meta)lenses with large spectral and angular bandwidth are potentially of immense importance for a wide range of applications, including high-performance smartphone imaging, optical projection, endoscopic optical imaging, and augmented/virtual reality. In this context, the bandwidth bounds derived in this paper may provide important information and insights toward the design of next-generation metalenses with, simultaneously, wide FOV, high resolution, and broad spectral bandwidth. 

\textit{Acknowledgment --} We acknowledge support from the National Science Foundation (NSF) with Grant No. 1741694, and the Air Force Office of Scientific Research with Grant No. FA9550-19-1-0043.



%


\bibliography{bib1}

\end{document}